\documentclass[
 reprint,
 amsmath,amssymb,
 prl,
 aps,
]{revtex4-2}

\usepackage{graphicx}
\usepackage{dcolumn}
\usepackage{bm}
\usepackage{multirow}

\newcommand{\iz}{{I}_{z}}
\newcommand{\ix}{{I}_{x}}
\newcommand{\iy}{{I}_{y}}
\newcommand{\ibz}{{I}_{\ib}}

\newcommand{\sz}{{\bf \sigma}_{z}}

\newcommand{\sy}{{\bf \sigma}_{y}}
\newcommand{\ib}{{\bf H}}
\newcommand{\e}{{\rm e}}
\newcommand{\I}{{\bf I}}
\newcommand{\w}{{\omega}}
\newcommand{\W}{{\Omega}}
\newcommand{\vph}{{\varphi}}
\newcommand{\vth}{{\vartheta}}

\newcommand{\ih}{{i\hbar}}

\newcommand{\Ha}{{\mathcal{H}}}

\newcommand{\al}{\alpha}
\newcommand{\be}{\beta}

\newcommand{\rb}[1]{\left(#1\right)}

\newcommand{\cb}[1]{\left\{#1\right\}}

\begin{document}

\preprint{}

\title{Kinematic Modulation in Driven Spin Resonance}

\author{Sunghyun Kim}
 \altaffiliation[Also at ]{University of Maryland}
 \email{sunghyun@umd.edu}
\affiliation{Laboratory for Physical Sciences, College Park, Maryland}%

\date{\today}

\begin{abstract}
The transition probability of a spin driven by a rotating magnetic field is reformulated. This work shows that, once projection onto the measurement basis is properly accounted for, the laboratory-measured probability is governed by both intrinsic spin dynamics and the time dependence of the measurement basis. For the rotating-field eigenbasis, this yields an additional kinematic modulation, leading to measurable deviations under strong driving. A unified probability expression is derived that subsumes the classic 1937 and 1954 formulations as limiting cases, while correcting the conventional treatment of magnetic resonance transitions.
\end{abstract}

\maketitle



Spin resonance underlies magnetic resonance spectroscopy, coherent spin manipulation, and modern quantum control. Conventionally, transition probabilities in rotating fields are evaluated by projecting the evolved state onto a measurement basis specified in the laboratory frame. This procedure has been widely successful, particularly in the weak-driving regime, where it reproduces the standard Rabi response~\cite{rabi37,sch37,rrs54,ramsey,gottfried}. Here the measurement basis denotes the states onto which probabilities are projected, while the observation frame denotes the reference frame in which that basis is specified. However, when the observation frame is time dependent, the measured probability need not be determined solely by the intrinsic spin dynamics.

This paper shows that the transition probability contains distinct dynamical and kinematic contributions. The latter arises from the motion of the observation frame, equivalently from the induced motion of the measurement basis. It becomes finite under strong driving, where it produces suppressed transition amplitudes and secondary oscillations. In the special case $\omega=\omega_0=\omega_1$, a second resonance appears with a $\sin^4(\omega\tau/2)$ form. A unified probability expression is derived that subsumes the classic 1937 and 1954 formulations as limiting cases, while correcting the conventional treatment of magnetic resonance transitions.



\section{\label{sec:inconsistency}Two Transition Probabilities}

The 1937 formulation of driven spin transitions~\cite{rabi37,sch37} and the 1954 rotating-frame formulation~\cite{rrs54,ramsey} yield different transition probabilities for a spin-$\tfrac{1}{2}$ system in a rotating magnetic field. The earlier result is
\begin{equation}\label{eq:1937}
    W_{1937}\!\left(-\tfrac{1}{2},\tfrac{1}{2}\right)
    = \left(\tfrac{\w\,\w_1}{\overline{\w}\,\W}\right)^2
    \sin^2\!\left(\tfrac{\W \tau}{2}\right),
\end{equation}
whereas the rotating-frame expression is
\begin{equation}\label{eq:1954}
    W_{1954}\!\left(-\tfrac{1}{2},\tfrac{1}{2}\right)
    = \left(\tfrac{\w_1}{\W}\right)^2
    \sin^2\!\left(\tfrac{\W \tau}{2}\right).
\end{equation}
The symbols are defined below. Despite their similar form, Eqs.~\eqref{eq:1937} and~\eqref{eq:1954} are not generally equivalent, because they correspond to different choices of observation frame and projection basis. 

\section{\label{sec:labframe}Spin Hamiltonian and Eigenbasis in the Fixed Laboratory Frame}
In this section, the spin Hamiltonian is defined in the laboratory frame. A spin $\I=(\ix,\,\iy,\,\iz)$ in a Cartesian coordinate system, is subject to a magnetic field $\ib(t)$ of amplitude $H$ and angles $\vph$ and $\vth$,
\begin{equation}\label{eq:ib}
    \ib 
    =H(\cos{\vph}\sin{\vth},\, \sin{\vph}\sin{\vth},\, \cos{\vth}).
\end{equation}
The Hamiltonian describing the spin--field interaction is
\begin{equation}
    \Ha = -\gamma\,\I\cdot\ib,
\end{equation}
where $\gamma$ is the gyromagnetic ratio.

Under the rotating-wave approximation (RWA), the field angles are specified as
\begin{equation}
    \vph=-\w t,\quad 
    \vth=\mathrm{const.},
\end{equation}
where $\w$ is the rotation frequency. The effective Larmor frequency is defined as $\overline{\w}=\gamma H$, and the characteristic frequencies are
\begin{equation}
    \w_0=\overline{\w}\cos{\vth}, \quad
    \w_1=\overline{\w}\sin{\vth}.
\end{equation}
Hence,
\begin{equation}\label{hamiltonian1}
    \Ha(t)
    = -\big(
        \iz\w_0
        + \ix\w_1\cos{\w t}
        - \iy\w_1\sin{\w t}
      \big).
\end{equation}

In the laboratory frame, denoted $(0,0)$, the state vector is expanded as
\begin{equation}
    \Psi(t)=\sum_m C_m(t)\,|m(t)\rangle_{(0,0)},
\end{equation}
and satisfies the equation
\begin{equation}\label{eq:SchrodingerEquation}
    i\hbar\frac{\partial \Psi}{\partial t} = \Ha \Psi .
\end{equation}
At each instant, the Hamiltonian defines a quantization operator $I_{\ib}(t)$ satisfying $[\Ha,I_{\ib}]=0$. It is given by the projection of the spin along the direction of $\ib$,
\begin{equation}\label{eq:quantizationframe}
    I_{\ib}(t) = \frac{\I\cdot\ib(t)}{H}
    = \frac{\iz\w_0+\ix\w_1\cos\!{\w t}-\iy\w_1\sin\!{\w t}}{\overline{\w}}.
\end{equation} 
Its eigenstates $\{|m(t)\rangle_{(0,0)}\}$ define the eigenbasis of $\Ha$ in the fixed laboratory frame, such that
\begin{equation}\label{eq:ih}
    I_{\ib}\,|m(t)\rangle_{(0,0)} = m\hbar \,|m(t)\rangle_{(0,0)},
\end{equation}
and 
\begin{equation}\label{Energy1}
    \Ha|m(t)\rangle_{(0,0)} = -m\hbar\overline{\w}|m(t)\rangle_{(0,0)}.
\end{equation}

\section{\label{sec:transformation}Unitary Transformations and \\the Wavefunction}

Equation~\eqref{eq:quantizationframe} shows that, in general, $I_z$ does not coincide with $I_{\ib}$ and therefore does not define the quantization axis~\cite{sch37}. However, by rotating the observation frame from $(0,0)$ to $(\alpha,\beta)$, where $\alpha$ and $\beta$ denote rotations about the $z$ and $y$ axes, respectively, $I_z$ can be aligned with the quantization axis and thus serves as the quantization operator~\cite{klk24}.

A rotation about the $z$ axis is implemented by a unitary transformation applied to Eq.~\eqref{eq:SchrodingerEquation}, generated by
\(
   \e^{\frac{\iz\al}{\ih}},\
   \e^{-\frac{\iz\al}{\ih}}
\)~\cite{sch37,lkk22},
\begin{equation}
    \Psi_{(\al,0)}=\e^{-\frac{\iz\al}{\ih}}\Psi,
\end{equation}
which represents a transformation to a rotating reference frame.
Hence, in the rotating frame $(\al,0)$,
\begin{equation}\label{eq:energyshift}
    i\hbar \frac{\partial \Psi_{(\al,0)}}{\partial t}
    =
    \left(
        \e^{-\frac{\iz\al}{\ih}}\Ha \e^{\frac{\iz\al}{\ih}}
        -\iz\dot{\al}
    \right)\Psi_{(\al,0)}
    =\Ha_{(\al,0)}\Psi_{(\al,0)}.
\end{equation}
In the frame co-rotating with $\ib$ in the $xy$ plane, $\al$ is specified as $-\w t$. $\Ha_{(\al,0)}$ becomes time-independent,
\begin{equation}\label{eq:hamiltonian2}
   \Ha_{(-\w t,0)}
   =-(\w_0-\w)\iz-\w_1\ix.
\end{equation}
Although $\Ha_{(-\w t,0)}$ is widely used for its simplicity~\cite{gottfried,ramsey}, Eq.~\eqref{eq:hamiltonian2} contains an additional term $-\iz \dot{\alpha} = \iz \w$ originating from the time-dependent frame rotation.

A second rotation
\(
    \e^{\frac{\iy\be}{\ih}},
    \e^{-\frac{\iy\be}{\ih}}
\),
\begin{equation}
   \Psi_{(-\w t,\be)}= \e^{-\frac{\iy\be}{\ih}} \e^{\frac{\iz\w t}{\ih}}\Psi.
\end{equation}
The angle $\be=\Theta$ is chosen such that
\(
    \tan\!{\Theta}=\left(\tfrac{\w_1}{\w_0-\w}\right),
\)
which diagonalizes the Hamiltonian
\begin{equation}\label{eq:hamiltonian3}
    \Ha_{(-\w t,\Theta)}= \e^{-\frac{\iy\Theta}{\ih}}\Ha_{(-\w t,0)} \e^{\frac{\iy\Theta}{\ih}}=-\iz\W,
\end{equation}
with the Rabi frequency $\W=\sqrt{(\w_0-\w)^2+\w_1^2}$. This second rotation introduces no energy shift, as $\Theta$ is constant. Hence, 
\begin{equation}
     \ih \frac{\partial \Psi_{(-\w t,\Theta)}}{\partial t} =
     \Ha_{(-\w t,\Theta)}\Psi_{(-\w t,\Theta)}.
\end{equation}

As seen from Eq.~\eqref{eq:hamiltonian3}, the Hamiltonian is simplified by two successive transformations, allowing the spin–field evolution to be readily described in the spin dynamical frame $(-\w t, \Theta)$ over the time interval from $t_1$ to $t_2$. 
Because this frame is already rotated relative to the laboratory frame by $(-\w t,\Theta)$ at the initial time $t_1$, the evolved state must be transformed back to the laboratory frame $(0,0)$ at the final time $t_2$. The resulting spin state is therefore~\cite{lkk22}
\begin{equation}\label{eq:labsolutioncompact}
    \Psi(t_2)
    =
    \e^{-\frac{\iz\w t_2}{\ih}}\e^{\frac{\iy\Theta}{\ih}}
    \e^{-\frac{\iz\W (t_2-t_1)}{\ih}}
    \e^{-\frac{\iy\Theta}{\ih}}\e^{\frac{\iz\w t_1}{\ih}}\Psi(t_1).
\end{equation}

\section{\label{sec:rotatedframe}Perturbative Nature of \\the Laboratory Basis}

Equation~\eqref{eq:labsolutioncompact} represents the time evolution of the spin state in the fixed laboratory frame $(0,0)$. 
The transition probability is then constructed by projecting the laboratory-frame solution of the Schr\"odinger equation onto this basis $|m\rangle_{(0,0)}$, yielding
\begin{equation}\label{eq:labtransition}
    \begin{aligned}
        &W_{(0,0)}(m,m')
        =\\~&
        \Big|\,
        _{(0,0)}\langle m|\,
        \e^{-\frac{\iz\w t_2}{\ih}}
        \!\e^{\frac{\iy\Theta}{\ih}}
        \!\e^{-\frac{\iz\W (t_2-t_1)}{\ih}}
        \!\e^{-\frac{\iy\Theta}{\ih}}
        \!\e^{\frac{\iz\w t_1}{\ih}}
        \,|m'\rangle_{(0,0)}
        \Big|^2 .
    \end{aligned}
\end{equation}

Conventionally, the measurement bases $\{|m\rangle_{(0,0)}\}$ defined in this frame are implicitly adopted as the quantization basis~\cite{rrs54,ramsey,gottfried}. With this identification, $\iz|m\rangle_{(0,0)}=m\hbar|m\rangle_{(0,0)}$, so the outermost exponentials contribute only phases and cancel in the modulus squared.

In the simple case of $I=\tfrac{1}{2}$, the spin operators can be represented by the Pauli matrices $\sigma_i$: $I_i=\hbar\sigma_i/2$. Since $\sigma_i^2={\bf 1}$, each exponential in~\eqref{eq:labtransition} reduces to a linear combination of ${\bf 1}$ and $\sigma_i$ with trigonometric coefficients. 
Hence, 
\begin{equation}\label{rrs}
    W_{1954}\!\left(-\tfrac{1}{2},\tfrac{1}{2}\right)
    =
    \sin^2\Theta\,
    \sin^2\!\rb{\tfrac{\W\tau}{2}}
    =
    \rb{\tfrac{\w_1}{\W}}^2
    \sin^2\!\rb{\tfrac{\W\tau}{2}},
\end{equation}
with the evolving time $\tau=t_2-t_1$.

The Hamiltonian and the corresponding Schr\"odinger equation, as well as their solution Eq.~\eqref{eq:labsolutioncompact}, are all formulated and evaluated in the laboratory frame. However, Eq.~\eqref{rrs} relies on the implicit assumption that the quantization axis is defined by $I_z$. 
As already indicated by Eq.~\eqref{eq:ih}, $I_z$ is distinct from the quantization operator $I_{\ib}$ associated with \(\ib(t) = H(\cos{\w t}\sin{\vth},\, -\sin{\w t}\sin{\vth},\, \cos{\vth})\). The conventional transition probability therefore corresponds to a perturbative description based on the bare Hamiltonian defined by the homogeneous field $\ib_0 = (0,\,0,\,H\!\cos\vth)$ and its associated quantization basis~\cite{rrs54}. 

Consequently, this misalignment removes a kinematic contribution associated with the observation frame. While this contribution appears only as a phase in the laboratory basis, it becomes physically consequential when the transverse driving field is not perturbatively small. 

\section{\label{sec:quantization} Rotating Observation Frame and Omitted Kinematics}

Equation~\eqref{eq:quantizationframe} shows that $I_z$ does not coincide with $I_{\ib}$ and therefore does not define the physical quantization axis~\cite{sch37}. 
In the presence of a rotating field, the Hamiltonian is proportional to the projection of the spin along $\ib(t)$, so that the energy eigenbasis is determined by $I_{\ib}$ rather than $I_z$. 
The operator $\iz$ defines the quantization axis in the frame that follows $\ib(t)$, namely the rotating-field frame $(-\w t,\vth)$: 
\begin{equation}\label{eq:rotatingket}
    |m\rangle_{(-\w t,\vth)} = \e^{-\frac{\iy\vth}{\ih}}\e^{\frac{\iz\w t}{\ih}}|m\rangle_{(0,0)}.
\end{equation}
Hence,
\begin{equation}\label{eq:quantization2}
    \e^{-\frac{\iy\vth}{\ih}}\e^{\frac{\iz\w t}{\ih}}
    \,\ibz(t)\,
    \e^{-\frac{\iz\w t}{\ih}}\e^{\frac{\iy\vth}{\ih}}
    =
    \iz.
\end{equation}

Although the rotating-field frame $(-\w t,\vth)$ restores $\iz$ as the quantization operator, Eq.~\eqref{eq:hamiltonian3} shows that this assignment is not unique. In the spin dynamical frame $(-\w t,\Theta)$, the effective Hamiltonian is likewise proportional to $\iz$, such that $[\Ha_{(-\w t,\Theta)},\iz]=0$. The same operator $\iz$ therefore defines the quantization axis in two distinct reference frames, reflecting the coexistence of kinematic and dynamical descriptions.

In the 1937 formulation~\cite{sch37, rabi37, guttinger}, the Hamiltonian is analyzed in this rotating-field frame:
\begin{equation}
    \Ha_{(-\w t,\vth)}
    = \e^{\frac{\iy\Gamma}{\ih}} \Ha_{(-\w t, \Theta)} \e^{-\frac{\iy\Gamma}{\ih}}
    =
     \e^{\frac{\iy\Gamma}{\ih}} \left(-\iz\W\right) \e^{-\frac{\iy\Gamma}{\ih}},
\end{equation}
where $\Gamma=\Theta-\vth$ is defined by the field geometry. The evolution observed in the rotating-field frame is
\begin{equation}\label{eq:rotatingfieldcompact}
    \Psi\!_{(-\w t,\vth)}(t_2)
    =
     \e^{\frac{\iy\Gamma}{\ih}} \,
    \e^{-\frac{\iz\W (t_2-t_1)}{\ih}}
    \,e^{-\frac{\iy\Gamma}{\ih}}
    \Psi\!_{(-\w t,\vth)}(t_1).
\end{equation}
In the $I=\tfrac{1}{2}$ case, the transition probability projected onto the rotating-field eigenbasis $|m\rangle_{(-\w t, \vth)}$ becomes
\begin{equation}\label{schwinger}
    \begin{aligned}
    &W_{1937}(m,m')
    =\\&~
    \Big|\,
    _{(-\w t,\vth)}\langle m|
    \e^{-\frac{i\sy\Gamma}{2}}
    \e^{\frac{i\sz\W (t_2-t_1)}{2}}
    \e^{\frac{i\sy\Gamma}{2}}
    |m'\rangle_{(-\w t,\vth)}
    \Big|^2,
    \end{aligned}
\end{equation}
yielding the 1937 result~\cite{rabi37,sch37},
\begin{equation}\label{wellEvaluation}
    W_{1937}\!\left(-\tfrac{1}{2},\tfrac{1}{2}\right)
    =
    \sin^2\Gamma\,
    \sin^2\!\left(\tfrac{\W\tau}{2}\right)
    =
    \left(\tfrac{\w\w_1}{\overline{\w}\W}\right)^2
    \sin^2\!\left(\tfrac{\W\tau}{2}\right).
\end{equation}

Accordingly, the rotating-field frame $(-\w t,\vth)$ restores $\iz$ in the sense of Eq.~\eqref{eq:quantization2}. Equation~\eqref{eq:rotatingfieldcompact} therefore provides an exact description of the spin dynamics in the frame co-rotating with the field. 
However, the transition probability is actually measured in the laboratory frame, and the rotation of the observation frame is not incorporated into the measured quantity. Equation~\eqref{schwinger} thus correctly captures the spin dynamics, but omits the kinematic rotation of the observer in the transition probability.

\section{\label{sec:hidden}Transition Probability in\\ the Dual Reference Structure}

Equations~\eqref{rrs} and~\eqref{schwinger} do not provide a complete description of the transition probability and its underlying structure~\cite{sch37,rabi37,rrs54}. Each expression captures only one aspect of the spin dynamics when a spin evolving under the time-dependent interaction $\Ha(t)$ is observed in a fixed frame. This limitation is reflected in the distinct constructions of the two results. The 1954 expression is obtained by simplifying the laboratory-frame solution in Eq.~\eqref{eq:labsolutioncompact} and projecting it onto the laboratory eigenbasis $|m\rangle_{(0,0)}$, whereas the 1937 expression is obtained by describing the same dynamics in the rotating-field frame $(-\w t,\vth)$.  These two prescriptions correspond to distinct reference frames for the quantum state. Their separation defines a dual reference structure, in which the observed transition probability reflects both the motion of the observation frame and the intrinsic spin dynamics.

To expose this structure, the laboratory basis is expressed in the rotating-field basis using Eq.~\eqref{eq:rotatingket},
\begin{equation}
    |m\rangle_{(0,0)}
    =
    \e^{-\frac{\iz\w t}{\ih}}
    \e^{\frac{\iy\vth}{\ih}}
    |m\rangle_{(-\w t,\vth)} .
\end{equation}
Substitution into Eq.~(\ref{eq:labtransition}) yields
\begin{equation}\label{realW}
    \begin{aligned}
        W_{(-\w t,\vth)}(m,m')
        =
        \Big|\,
        _{(-\w t,\vth)}\langle m|
        \e^{\frac{\iy\Gamma}{\ih}}
        \e^{-\frac{\iz\W\tau}{\ih}}
        \e^{-\frac{\iy\Gamma}{\ih}}
        \\
        \hspace{40pt}\times
        \e^{-\frac{\iy\vth}{\ih}}
        \e^{-\frac{\iz\w\tau}{\ih}}
        \e^{\frac{\iy\vth}{\ih}}
        |m'\rangle_{(-\w t,\vth)}
        \Big|^2 ,
    \end{aligned}
\end{equation}
with time interval $\tau = t_2-t_1$. Equation~\eqref{realW} separates into two distinct rotational contributions: a dynamical rotation generated by the effective Hamiltonian with frequency $\W$, and a kinematic rotation at frequency $\w$ associated with the motion of the observation frame. Accordingly, the laboratory-frame evolution is not purely intrinsic, but reflects both spin dynamics and frame motion.

For the $I=\tfrac{1}{2}$ system, the combined evolution operator yields
\begin{equation}\label{finalW}
    \begin{aligned}
        &W\!\left(-\tfrac{1}{2},\tfrac{1}{2}\right)
        =
        \rb{\tfrac{\w_1}{\W}}^2
        \sin^2\!\rb{\tfrac{\W\tau}{2}}
        \sin^2\!\rb{\tfrac{\w\tau}{2}}\\
        &\quad+
        \cb{
            \rb{\tfrac{\w\w_1}{\W\overline{\w}}}
            \sin\!\rb{\tfrac{\W\tau}{2}}
            \cos\!\rb{\tfrac{\w\tau}{2}}
            -\rb{\tfrac{\w_1}{\overline{\w}}}
            \cos\!\rb{\tfrac{\W\tau}{2}}
            \sin\!\rb{\tfrac{\w\tau}{2}}
        }^2 .
    \end{aligned}
\end{equation}

Equation~\eqref{finalW} gives the complete transition probability when both reference frames are treated explicitly. Neglecting the kinematic rotation recovers $W_{1937}$, while projection onto the laboratory eigenbasis yields $W_{1954}$. The transition probability is therefore not intrinsic to the spin system, but a relational quantity defined between distinct reference frames.

\section{\label{sec:closing1}Resonances}

At exact resonance, $\w=\w_0$, transition probability takes the form
\begin{equation}\label{eq:atresonance}
    \begin{aligned}
    &W_{\text{res.}}\!\left(-\tfrac{1}{2},\tfrac{1}{2}\right)
    =
    \sin^2\!\rb{\tfrac{\W\tau}{2}}
    \sin^2\!\rb{\tfrac{\w\tau}{2}}\\
    &\!
    +\cb{
    \rb{\tfrac{\w}{\overline{\w}}}
    \sin\!\rb{\tfrac{\W\tau}{2}}
    \cos\!\rb{\tfrac{\w\tau}{2}}
    -
    \rb{\tfrac{\W}{\overline{\w}}}
    \cos\!\rb{\tfrac{\W\tau}{2}}
    \sin\!\rb{\tfrac{\w\tau}{2}}
    }^2 .
    \end{aligned}
\end{equation}
Equation~\eqref{eq:atresonance} shows the coexistence of dynamical oscillations $\W$ and kinematic modulation $\w$.

At the weak-driving limit $\omega_1\ll\omega_0$, the kinematic contribution vanishes and the transition probability reduces to the conventional Rabi form,
\begin{equation}
    W_{\text{weak res.}}\!\left(-\tfrac{1}{2},\tfrac{1}{2}\right)
    =
    \sin^2\!\rb{\tfrac{\W\tau}{2}}.
\end{equation}
Under strong driving $\omega_1\sim\omega_0$, the kinematic rotation introduces an additional modulation that suppresses the maximum transition probability and produces secondary oscillations. This effect is intrinsic to the dual-reference structure and does not originate from counter-rotating-field corrections such as the Bloch--Siegert shift~\cite{bloch40}.

Notably, a second resonance occurs at $\w=\w_0=\w_1$, where the transition probability reduces to
\begin{equation}
      W_{\text{2nd res.}}\!\left(-\tfrac{1}{2},\tfrac{1}{2}\right)
    =
    \sin^4\!\rb{\tfrac{\W\tau}{2}}=
    \sin^4\!\rb{\tfrac{\w\tau}{2}}.
\end{equation}

\nocite{*}
\bibliography{reference}

\end{document}